# U-Note: Capture the Class and Access it Everywhere

Sylvain Malacria [1], Thomas Pietrzak[1,2], Aurélien Tabard[1,3], Éric Lecolinet[1]

[1] Telecom ParisTech – CNRS LTCI UMR 5141, 46 rue Barrault, 75013, Paris, France
{malacria, elc}@telecom-paristech.fr
[2] University of Toronto, 40 St. George Street, Toronto, Ontario, Canada
pietrzak@dgp.toronto.edu
[3] IT University of Copenhagen, Rued Langgaards Vej 7, DK-2300 Copenhagen, Denmark
auta@itu.dk

**Abstract.** We present U-Note, an augmented teaching and learning system leveraging the advantages of paper while letting teachers and pupils benefit from the richness that digital media can bring to a lecture. U-Note provides automatic linking between the notes of the pupils' notebooks and various events that occurred during the class (such as opening digital documents, changing slides, writing text on an interactive whiteboard...). Pupils can thus explore their notes in conjunction with the digital documents that were presented by the teacher during the lesson. Additionally, they can also listen to what the teacher was saying when a given note was written. Finally, they can add their own comments and documents to their notebooks to extend their lecture notes.
We interviewed teachers and deployed questionnaires to identify both teachers and pupils' habits: most of the teachers use (or would like to use) digital documents in their lectures but have problems in sharing these resources with their pupils. The results of this study also show that paper remains the primary medium used for knowledge keeping, sharing and editing by the pupils.
Based on these observations, we designed U-Note, which is built on three modules. U-Teach captures the context of the class: audio recordings, the whiteboard contents, together with the web pages, videos and slideshows displayed during the lesson. U-Study binds pupils' paper notes (taken with an Anoto digital pen) with the data coming from U-Teach and lets pupils access the class materials at home, through their notebooks. U-Move lets pupils browse lecture materials on their smartphone when they are not in front of a computer.

## 1 Introduction

Many teachers are now comfortable with digital media. Additionally, digital equipment such as PCs, video projectors, interactive white boards (IWB), etc., have become increasingly affordable. This makes it possible to use digital material in the classroom, not only at the university level, but also in middle and high schools. Laptops and mobile devices are now widespread, so that most pupils can work with digital documents at home, at the library, at other pupils' places and even in public transportation.

Various studies of augmented classrooms have already been published [1,4,12,14] with a focus on colleges and universities. In this paper, we present U-Note, a system designed for middle and high schools. The needs and the requirements of universities are different from those of high schools. From our preliminary interviews, we observed the importance of handwritten notes. Pen and paper are still the main tools used by pupils in class for several good reasons, pen and paper are cheap, flexible, easy to use, and not distractive [13,15]. But, back at home when the pupils read what they wrote in their notebooks, they cannot easily access the digital media presented during the class. They also don't have any way or reviewing the teacher's oral explanations they may have missed. Finally, while pupils now commonly use computers at home and elsewhere, there is no simple way to link their digital work (for example, searching and reading web pages) with their notebooks, which remain their main means for storing, organizing and retrieving information.

We designed U-Note by taking these findings into account. U-Note aims at linking the pupils' handwritten notes with the material that was presented during the class, with a high level of granularity. The pupils' notebooks serve as a means for referencing and accessing of digital media, oral explanations and the writing on the whiteboard, hence providing a physical medium for retrieving information from various sources. The notebooks provide a link between the pupils' works in class, at home and at any other locations. Furthermore, notebooks are designed for active reading so that pupils can enrich their personal libraries by adding their own digital documents. In all cases U-Note allows fine-grained correspondence. For instance, a phrase, symbol, or drawing can be linked with a simple slide of a presentation or an excerpt located at a specific location in a web page.

This paper begins with a presentation of existing annotation and note taking systems. Next we present interviews with elementary, middle and high school teachers. This stage helped us to refine our goals and to focus on our users' needs. We then describe U-Note and the features that appeared to be useful according to our investigations. Finally we conclude and present future work.

## 2 Related Work

### 2.1 Presentation Tools

Various systems allow subsequent access to captured live experiences. With Ubiquitous Presenter [22], a classroom presentation tool, the instructor can annotate slides with a Tablet PC while showing them and giving the lecture. The students can view the live presentation with narration and digital ink using standard PCs. The captured presentation is saved to a web server and can be retrieved later by students as a video. Recap [8] enables users to capture the lecture with more details than Ubiquitous Presenter; the presentation is indexed by slide number and by the pen strokes. Students can access this capture after the class through an ActiveX enabled web browser. However, neither Ubiquitous Presenter nor Recap provides the capability to link the multimedia data shown in class with students' notes.

Classroom2000 [1], which later became eClass [4], is a classroom presentation tool that allows an instructor to annotate slides on an interactive whiteboard. These annotations are linked with a video and audio recording of the class, and with the web links opened during the lecture. A longitudinal evaluation of this tool showed the usefulness of the links between the documents and the audio recordings. It also underlined the fact that students took fewer notes when using the system, which is not surprising since the teacher provides his or her notes. The advantage of this is that students may concentrate on the material. However, some authors have argued that taking notes has an important role in the memorization process [7].

StuPad [19], which integrates a note taking system with pen-based video tablets, provides students with the ability to personalize the capture of the lecture experiences. However, such equipment is currently not suited to middle and high school where paper notebooks are still widely used. Besides, as demonstrated in [13,15], interfaces departing from classic GUIs such as pen tablets and graphical tablets tend to deteriorate performance, especially for low-performing students.

The Digital Lecture Halls (DLH) project focuses on large audiences and provides the lecturer with a tool to control his lecture through a dedicated interface on a pen-based tablet [12,14]. The lecturer can thus write on the digital blackboard through the tablet and annotate his presentation, while keeping eye-contact with the audience. The audience can use specific software on their digital devices (such as smartphones or laptops) to mark parts of a lecture as particularly interesting, or to ask a question. While these solutions make perfect sense for large and mature audience lectures, middle school and high school have a much smaller and co-located audience were contact with the pupils is easy. Furthermore pupils tend to be easily distracted and new equipment and software that interferes with the course may overly distract pupils.

These systems combine all captured data into a unique stream and broadcast it through a web interface or as downloadable videos. The students cannot open (or eventually edit) the documents with their usual tools nor they can benefit from the flexibility of paper for indexing or annotating what is displayed on the screen. Having a separate medium for annotating (the notebook), which does not consume space on the screen is also another advantage, especially when using small laptops or mobile devices. Finally, another important requirement is the need for the students to link their own notes with the digital work they perform at home, a capability that is not supported by these systems (except StuPad, however with StuPad pupils can only attach a keyboard-typed text to a whole lecture, they cannot attach more sophisticated digital content such as web pages). Moreover, they cannot link this content precisely enough to link it to a specific sentence they may have written during the class.

### 2.2 Note Sharing and Annotation Tools

Miura *et al.* [11] present AirTransNote an interactive learning system that provides students with digital pens and PDAs. AirTransNote collects the handwritten drawings of the students and transmits them to the teacher's PC, so that teachers can closely monitor their students' work. A second version of AirTransNote [10] allows the teacher to replay the students' notes on a PC and to provide feedback on the PDAs of the students. These two works involved digital pens based on ultrasonic technology. When the student puts the pen down, ultrasonic waves are generated and provide the tip position relative to a sensor plugged at the top of the sheet of paper. However, the student has to specify when he starts to write on a new sheet of paper and cannot modify previously written pages. The third version of AirTransnote exploited the Anoto technology [2] as a way to avoid these limitations. The Anoto technology uses a small camera embedded in a ballpoint pen to read a dot-pattern printed on paper in order to locate the pen's position. Although Miura *et al.* investigated the various versions of AirTransNote during experimental lectures at a senior high school, their studies mainly focused on note sharing and real-time feedback for students during short tests in class.

Using CoScribe [16], the teacher starts the lesson by giving printouts of the slides to the students, who can then directly create handwritten annotation on the teacher's printouts using an Anoto pen. They also can structure and tag their annotations for later retrieval. Finally, they can collaborate with other students by sharing their annotations. However, contrary to the system we propose, CoScribe uses the printouts as a central media and does not provide a way to associate the teacher's material with the notes in the student's notebooks.

### 2.3 Augmented Notebooks

The Audio Notebook [17] is a device combining a paper notebook with a graphical tablet and an audio recorder. The user can then listen to what was recorded when a specific note was written just by tapping on it. The Livescribe digital pens [9] extend this idea by including the audio recorder within an Anoto digital pen, making it possible to get rid of cumbersome devices such as the graphical tablet. However, these systems are limited to audio recording and cannot link handwritten notes with other types of digital data.

Other studies generally based on the Anoto technology have been devoted to augmented notebooks. Brandl et al. designed NiCEBook [3], an augmented notebook that enhances natural note taking. NiCEBook provides tagging functionality and allows users to share their handwritten notes in a vector format via e-mail. However, this system is not intended to link digital documents with personal notes.

Yeh et al. developed a notebook for field biologists [23] that associates handwritten notes with GPS coordinates, photos they shot or samples they found in the field. West et al. designed a similar system for scrapbooking [20]. Their system allows combining handwritten notes with media documents such as photos, videos and sounds using explicit gestures. Finally, Tabard et al. proposed Prism [18], a hybrid notebook that aggregates streams of digital resources (documents, web pages, emails) with biologists' notebooks. Its long-term deployment showed that, among all the data streams that were aggregated, users tend to rely on one of them as their master reference (which was generally the paper notebook). All these studies focused on different contexts than the electronic classroom. While they share some similarities with our work (as they also rely on augmented notebooks), the requirements of our application domain are different. For instance, our system allows fine-grained correspondence between handwritten notes and specific locations in digital documents, a feature that was not needed in these previous systems. The ability to link and further access the digital events that occurred during the class, and to enrich and personalize this data later at home and in other contexts, constitutes important improvements over these previous technologies, in our high school teaching domain.

## 3   Motivations and Interviews

To better understand how French teachers and pupils currently use digital material in the classrooms, we visited a school located in inner Paris. Based on insights from interviews with the teachers, we developed three online questionnaires that we distributed to teachers and pupils.

### 3.1 Method

**Interviews.** We interviewed three teachers with pupils from middle school (11-15 years old) and high school (15-18 years old). Each teacher was interviewed separately for one hour. We focused on their use of paper and digital materials during 'normal' classes, practical classes, and outside of the classroom.

**Questionnaires.** We gathered information about the uses and needs of paper and new technologies by means of three questionnaires [25]: two for the teachers and one for the pupils. We asked one elementary and two high school teachers to give us feedback on preliminary versions of the teachers' questionnaire. This helped us to rephrase some questions so that they would better match the learning practices. We asked teachers of several schools to complete the questionnaire online. Eighteen teachers (15 female, 3 male) answered the questions (5 in elementary school, 8 in middle school and 5 in high school). The elementary school teachers each taught multiple topics. The other teachers either taught Mathematics (5), Literature (3), English (2), History and Geography (1), Physics and Chemistry (1) or Economy and Management (1).

We then designed a second teachers' questionnaire and a pupils' questionnaire to confirm and complete the answers of the first questionnaire. Nine pupils answered the pupils' questionnaire, 4 from middle school and 5 from high school. Twelve teachers (9 female, 3 male) answered the second teachers' questionnaire (3 were teaching in elementary school, 5 in middle school and 4 in high school). As before, elementary school teachers each taught several topics while the other teachers either taught Mathematics (3), Foreign languages (3), Literature (2) or History and Geography (1).

### 3.2 Results

We identified whiteboards, books, and paper handouts, as the teachers' main resources for knowledge keeping, sharing or editing. Pupils mainly relied on notes written on paper and handouts to record the lectures and learn their lessons.

Teachers used multimedia equipment such as computers and video projectors as a way to augment existing lectures with digital documents, but not systematically for all lessons. These digital materials are of different kinds, depending on the topic of the lesson: for instance audio materials in language classes, videos in history or biology classes, and interactive demonstrations in mathematics or biology classes. A major problem we identified is that these materials are not often available to pupils after the lecture. These modern digital materials cannot be printed out and distributed as hardcopy, they must be sent by email or posted on an online teaching portal. However this does not appear to be a widespread practice, except for teachers with technical skills, who generally preferred to put digital materials on their personal web sites. Finally, we also found that elementary school teachers scarcely use digital materials compared to middle and high school teachers.

Next we summarize the results of the three questionnaires and analyze the most interesting points.

**Teaching Materials.** Not all of the teachers have easy access to multimedia equipment such as computers and video projectors. For instance 3 out of 12 reported difficulties in having access to a video projector connected to a computer as often as they would like, and 4 out of 12 (at the elementary school) do not have access at all (Fig. 1, left). Practical constraints, such as the scarcity of, and time needed for installing these devices, reduced their availability, although the teachers showed interested in using multimedia equipment.

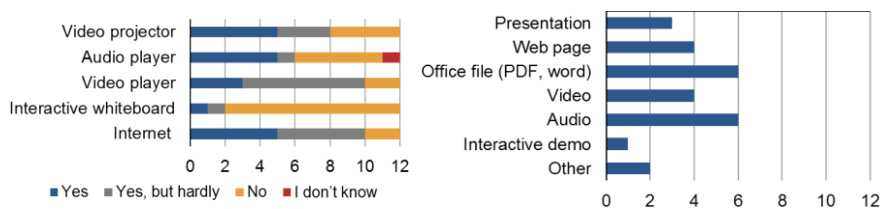

**Fig. 1.** Left: Number of teachers having access to these devices at school.
Right: Number of teachers using a type of media during their lessons

The analysis of the type of the media used during lessons (Fig. 1, right) shows that these resources are not necessarily in digital format: audio (used by 6 out of 12 of the teachers) and video (4 out of 12) could also be broadcasted in class using analog devices (e.g., VCR players). Non-digital media formats would be a barrier to sharing such documents with pupils. Other documents, such as pictures, exercises or tests in PDF formats are widely used but often printed on paper and distributed to pupils rather than projected during the class.

Four out of twelve of the teachers use web pages, which are used regardless of the lecture subject (Mathematics, History or Literature), this shows the potential of this media. Also, web pages are easy to share and flexible: teachers can give links or printouts, and they can also contain video or audio files.

Using digital materials in the classroom is a rather new practice. Even if relatively few teachers used them commonly, most of them thought this was going to increase in the future. As one of the high school teachers said: "*Last year, I renewed my lectures, and used many more slideshows and videos. The answer would have been very*

*different two years ago: due to difficult access to the devices, I would not have bothered adapting or building my lectures around these resources.*"

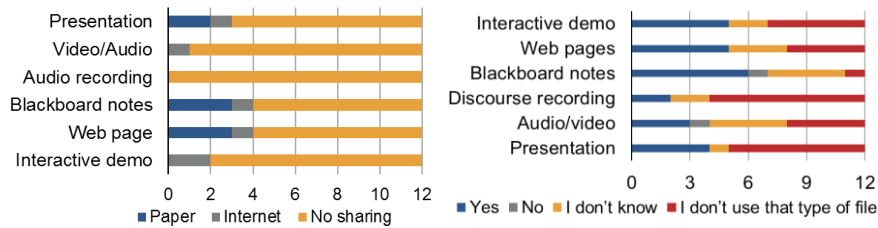

**Fig. 2.** Left: Do you share this type of file with your pupils? Right: Would you accept the creation of a history log of the files you use during your class?

**Sharing Resources.** We also observed that the sharing of materials was problematic. While half the teachers used video or audio resources, less than 10% sharing these resources after their lectures (Fig. 2, left). Teachers generally distributed their files as printouts that are distributed to pupils. However, this can affect the quality of the information. For example, audio transcriptions in foreign languages can be useful for practicing grammar and vocabulary, but is useless for practicing oral pronunciation. In addition, while some digital files like slideshows might be successfully conveyed on paper, other media like videos or interactive presentations would be difficult to convey through a printout if the material contains animation or other dynamic features.

Both pupils and teachers could take advantage of an easy sharing solution. Most teachers said they would accept using a system that would automatically transfer the digital files used during the lesson to the pupils (including a capture of the blackboard). But some teachers would only accept sharing files upon specific conditions such as the ability to control pupils' access and which files would be shared (typically, not the blackboard capture). As one teacher stated "*I'm in favor of transferring any type of digital data except the notes taken on the blackboard, to make sure of pupils take their own notes during the class*."

**Capture in the Classroom.** We further investigated whether teachers would agree to use a system that creates a log of the digital documents used during the lesson. As shown by figure 2 on the right, most teachers would accept that kind of system (given that they already used digital documents during their lectures) or do not know yet. In fact, less than 10% would refuse to use that kind of system for blackboard notes and audio/video files. Several teachers answered "I don't know", probably because of their concerns regarding the impact of sharing so much information with the pupils. As explained by one of the teachers, "*If everything is sent to pupils, they will not take notes anymore. Nevertheless, it can be useful for pupils to be able to access what happened in class to check a lesson from time to time or in case they missed a specific lecture*". Hence, as noted above, providing the ability for teachers to control pupils' access to files is a key factor for acceptance.

**Note Taking by Pupils.** Pupils write notes on notebooks in all classes. Teachers progressively teach note taking to pupils, from the first years of middle school to high school. Initially the teachers write on the whiteboard, then progressively move to dictation, writing only keywords on the whiteboard. By the end of high school, pupils create their notes from the teachers' speech without the teachers having to provide written text. Yet, the teachers adapt the way of speaking from one class to another. As they dictate, teachers make sure that the pupils are still following or will slow down, moving from writing only keywords to the whole course on the whiteboard as necessary. Several teachers stressed that they wanted to ensure that their pupils take notes. Hence, the principle of keeping the paper notebook as the central pupil's media, as we propose with U-Note, fits the teachers' recommendations.

Pupils not only take notes during lessons but also when working on computers in lab classes. For instance, during a visit to a high school, we observed pupils performing exercises on Open Office spreadsheets. They were asked to report results on paper printouts and explain how they solved the problem. Paper made it easy for the teacher not only to go through the pupils' work, but also to annotate the pupils' work and write comments and advice.

Without doubt, paper is still the most widely used media. As shown in figure 3 (left), pupils write in their notebooks on a daily basis and during most of the classes they attend. This observation was corroborated by the pupils' questionnaire answers. Two thirds of the teachers (12 out of 18) said that their pupils were writing on printouts everyday, and one third (6 out of 18) at least every week. This intensive use of handouts was not only explained by the good properties of paper (which is easy to use, to share, etc., as are notebooks) but also by the fact that all of the teachers we interviewed could easily access a photocopier (figure 1, left).

The amount of time spent in writing during the class is important (figure 3, right). According to teachers, most pupils spend more than 10 minutes writing in their notebooks during 55 minutes-long lessons (15 out of 18).

The use of paper was well summarized by one of the teachers: "*The notebook is the default medium. It is the only medium that lets us hope that pupils keep their materials from one class to the next one. I encourage pupils to write as much as possible on the handouts I give them so that they appropriate them, but it is somewhat difficult. They do not dare and feel reassured to write in their own notebook.*"

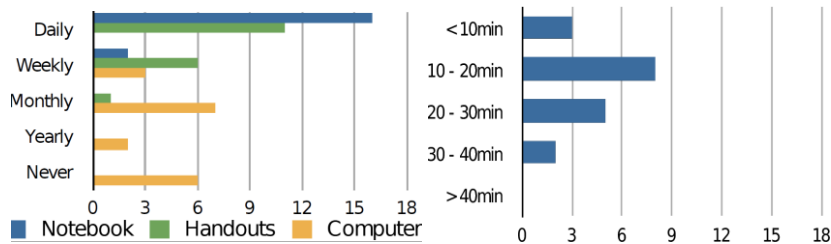

**Fig. 3**   Left: on which media do your pupils take their notes?
Right: how much time do your pupils write during one hour of class?

**Access to Notes.** All pupils claimed to be reading their lessons at home[1]. More interestingly, almost all of them (13 out of 15) declared reading their notes at school and about half (7 out of 15) declared reading their lessons while on public transportation. This underlines the importance of situations where the pupils cannot use their computer, and hence cannot access the teachers' materials.

### 3.3 Privacy Considerations

The remarks raised by the teachers pointed out some interesting concerns we initially overlooked, pertaining to who has access to the data. For example, many teachers felt that if the school administration could access their materials, the administration could also control how they work and what happened in the classroom. Teachers indicated that they wanted to keep the control of their own lectures. Some teachers also fear that these materials may be misused if given to the pupils. For example, some recordings could be posted on social networks to make fun of them (because of their accent, when they mumble or when they make mistakes). But one reluctant teacher said: "*If we can be assured that these recordings would only be used by pupils that want a new explanation of the lesson, I would definitely be for it. So, with strong guardrails, this could be interesting!*"

### 3.4 Implication for Design

This preliminary work indicated that paper is still the central medium for organizing information in the classroom. As we pointed out earlier, paper notes do not currently hold any type of digital information. In the following section, we present the system we proposed for augmenting paper notes with digital information. This system is based on the observations we made through the interviews and questionnaires presented above. On the teachers' side, the system captures events during the class, and on the pupils' side the events are linked to the pupils' notebooks.

---

[1] Even if the survey was anonymous, there may be some issues in trusting these numbers as the pupils answering the survey were likely the most dedicated ones.

On the one hand teachers need:

- Devices for playing digital materials during their lectures.
- Digital materials (personal or academic data provided by institutions or editors).
- Systems for making these materials easy to store, share, and access. Additionally, these systems must provide access control so that teachers can specify what should be freely available to the pupils.

On the other hand pupils need a simple means for accessing the data in various situations (at home, in the library, in public transportation) and for associating it with their own notes. Hence:

- The notebook should remain the central media for pupils, but enhanced to make it possible to retrieve all useful information.
- The links between the pupil's notes and the related digital media should be as specific and precise as possible in order to let pupils easily locate the information they are looking for.
- When pupils misunderstand some parts of the lesson they should be able to easily access the corresponding oral or written explanations, whenever possible.

## 4  Scenarios

Pupils use their notebook and the associated information in various locations (in the classroom, at home, etc.). Depending upon the situation, they may not have access to the same devices and materials. We describe below several typical situations we identified as relevant.

### 4.1 In the Classroom

Ms. Green is giving a lecture on animals' breathing mechanisms in her life-science class. She introduces the lecture by raising questions regarding the breathing of different animals on earth, and in the water and air. Pupils interact with her and take notes on the introduction she dictates. After the introduction, she projects a video she had prepared earlier on her laptop.

The video presents breathing organs from three different animals: cows, crickets and salmon. Between each animal, she pauses the video and dictates to pupils what they just observed. While explaining the video she also draws diagrams on the whiteboard that pupils copy. Ms. Green then questions the pupils, so that they can progressively annotate the diagrams with arrows and labels.

### 4.2 At Home

A few days later Johnny, one of Ms. Green pupils, is doing his homework for the next class. While taking his notebook, he plugs his ANOTO pen to the computer to sync paper and digital notes.

The first exercise consists in identifying the different breathing organs of a frog. The case is complex as frogs use both lungs and skin to breathe. After checking his manual, he goes back to his paper notes and taps with his pen on the diagrams he drew in his notebook to be sure he did not forget any cases. The associated digital materials appear on his computer. While looking at the diagram, he notices a link to the video presented in the class, and loads it.

As Johnny can easily be distracted, he spots a link related to frog breathing in the comments and clicks on it. This link leads to a web tutorial that is helpful for his exercise. While reading it he uses the web capture tool (provided by his U-Note browser) to save captures of the most interesting parts to his digital notebook. These captures, especially the diagrams they contain, will be useful later when studying for the test. They may also be useful for his friend Frank, who often calls for help.

### 4.3 In a Mobile Situation

At the end of the month, Johnny has to prepare for his test. During a one-hour break, he goes to the library to review his lessons. As he suspects that the frog case or a similar one could be asked, he loads the link he saved a few weeks ago dealing with frog breathing and goes through it again. But, unfortunately, Johnny does not have enough time to finish reviewing before the next class. This is not really a problem as he will be able to continue in the bus, after the class, when going back home.

## 5 U-Note

U-Note was designed and developed to interact with notes and digital documents in all of the situations described above. U-Note comprises three tools. U-Teach is the capture system used by the teacher. U-Move is the mobile client that can be used for browsing digital documents on a mobile phone. U-Study is the pupil software. It allows viewing and editing notes and the associated digital material on a PC. We describe these tools below through three tasks: capture, access, and annotation.

### 5.1 Capturing the Class

The classroom is the main capture location. The teacher provides information through speech, writing on the blackboard and digital documents. Meanwhile, pupils write their lesson in their notebook. Using a paper notebook is important for several reasons. Writing helps the student remember and understand [7] and using paper rather than computers prevents distraction [13]. Moreover, users' notes combined with audio recordings proved to be a powerful means of indexing meetings [21] (a situation similar to lectures). Furthermore, notes can serve as user-defined indexes for referencing events the user considers important [21].

The U-Teach module captures the information related to events occurring during the class while the pupils take notes. Teachers described in their interviews how they adapt the flow of their lessons to make sure that the pupils are still following. This ensures that the pupils' notes are synchronized with the teachers' discourse and the documents presented on the digital board. The U-Study module, described later, creates high granularity links between these events and the pupils' notes.

As noted earlier, teachers use multimedia files during their lectures, in particular in middle school and high school. The purpose of this part of the system is to record the context of the class. It is composed of several programs and plugins.

First we developed a PowerPoint extension that detects and records important events such as slide changes and the loading/unloading of presentation files. This plugin provides information regarding which slide of which presentation is shown at a given time. The same functionality is offered for web pages through a Firefox extension. As teachers often use audio or video recordings in their lectures, we also developed a dedicated multimedia player that logs actions such as load, unload, play and pause on these files. All these software components send events to a central server. This server generates a log file of the lecture that is accessible to the pupils' application. The current implementation does not check for access rights. While we did not focus on security issues, these could be resolved with a password system or certificates.

Finally we also capture the teacher's oral explanations using audio recording software that is running on the teacher's PC. The program allows the teacher to stop the recording, for instance if there is a disruption in the class. Additionally, the system can also take into account the events generated by an interactive whiteboard when such a device is available. The teacher's writing on the whiteboard is also made available to the pupils.

### 5.2 Accessing Digital Materials from the Notebook

Two tools are offered to pupils for accessing and enriching the information contained within the notes in their notebook: U-Study, a desktop application for working at home, and U-Move, a web application that provides limited but still useful functionality in mobile situations.

**U-Study.** The U-Study module is a desktop application that displays a copy of the pupil's handwritten notes and provides the digital documents used by the teacher during the lecture (Fig. 4). When reviewing a lesson, pupils may read any part they did not understand in class. By clicking on the corresponding notes in the notebook they can access the data related to this specific part of the course such as the oral recording at this specific moment, the slide, the web page or the video that was displayed at that time, and what the teacher was writing (depending on which media were used and captured during the class). Additionally, pupils can also open digital documents with their favorite applications, so that they can browse, and even edit and save them, more conveniently. We developed U-Study in Java with QT Jambi and used PaperToolkit [24] for retrieving the Anoto strokes from the digital pen.

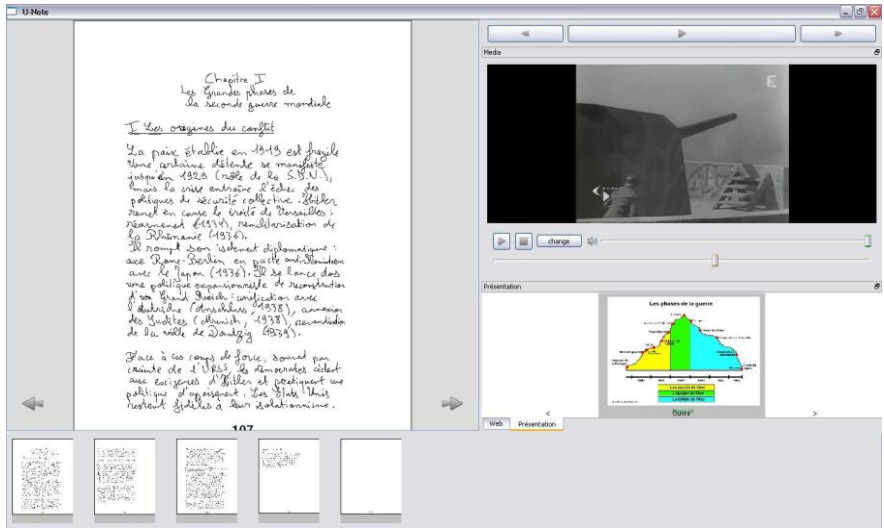

**Fig. 4.** Screenshot of the U-Study main view

*The notebook view*. The U-Study main window provides a view of the notebook (Fig. 5) that contains the strokes that were provided by the Anoto pen. The user can browse the pages of the notebook by clicking on two buttons. This view is mainly useful when the pupil's notebook is not at hand, it can be hidden otherwise.

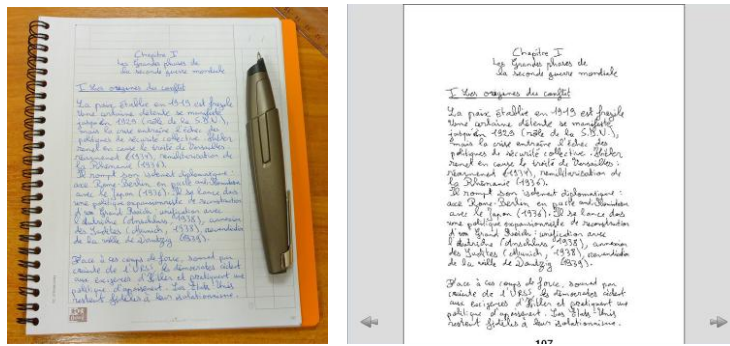

**Fig. 5.** Paper notebook and notebook view in U-Study. The pupil can use two buttons to browse the digital copy of his notebook.

*The miniature area*. U-Study can also be used to explore the teacher's documents while reading the notes. The "miniature area" can currently display four kinds of viewers (Fig. 6). The first viewer displays the miniature slides of a slideshow. It can be used for browsing the miniatures or for opening the original PowerPoint files. The second viewer allows the display of the web pages that were seen in class and to interact with dynamic content (hyperlinks, flash animations, etc.) when available. The third viewer is a multimedia player, which can play audio and video files. The fourth viewer is an interactive whiteboard viewer that displays the teacher's drawings and writings if a whiteboard was used during the class. Any viewer can be displayed or hidden on demand. The user's favorite applications can also be used when preferred.

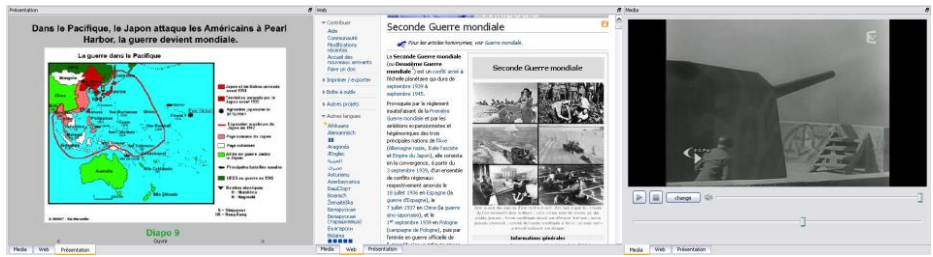

**Fig. 6.** Miniatures: PowerPoint, web pages and video.

*The thumbnails bar*. The thumbnail bar provides a visual link between the notebook and the digital documents (Fig. 7). U-Study displays a thumbnail for each page of the paper notebook. When the pupil moves the mouse cursor, contextual tool-tips pop-up. The tool-tips contain a thumbnail of the specific parts of the documents that were displayed while the pupil was writing the page. Typically, each slide, video sequence, and web page has a corresponding thumbnail. When the pupil clicks on a thumbnail, the miniature of the document pops out in the appropriate miniature widget (as described above). In order to save screen space, U-Study displays six thumbnails simultaneously. Buttons located on the sides of the tooltip provide access to next or previous thumbnails if more than six documents are associated with the current page.

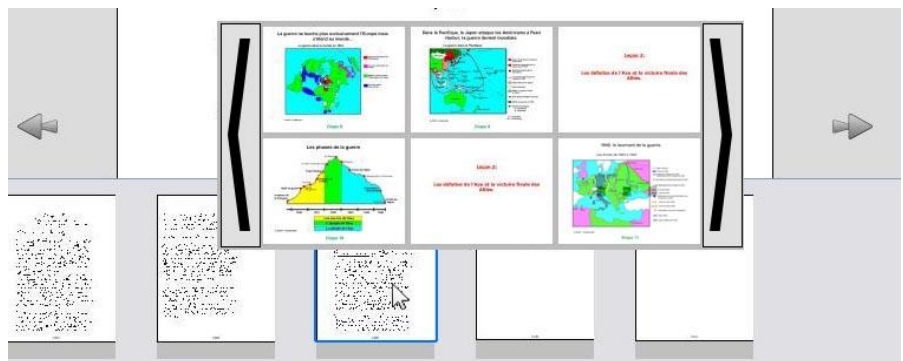

**Fig. 7.** Thumbnails bar.

*Replay*. When the dynamic of the class is important, the pupil can "replay the class" from a given point. A red dot moves over the handwritten strokes on the notebook view to show what was written and when. The miniatures are updated in the corresponding views to show which materials (and which specific subparts of them) were shown in the class at that time. The pupil can interactively control the replay speed and pause or resume at any moment during the replay. Besides the red dot that continuously shows the temporal position in the notebook, any handwritten strokes written at a later time can optionally be grayed out to enhance the visual feedback (this is configurable by the user, as this feature may decrease readability).

*Interaction with the notebook.* The pupil can start a replay by tapping on their notebook with the Anoto pen connected in streaming mode with a PC. The system identifies the timestamp of the stroke specified by the pupil and the red cursor moves to the same position in the notebook view. The replay resumes at this time and the digital materials shown in class are opened in the miniature area. The fact that the notebook serves as a link between all the lecture materials makes this feature especially useful. The internal clock of the digital pen is synchronized to the clock of the PC each time the pupil plugs it to its computer. To ensure synchronization, the clocks in the pupils' and teachers' computers have to be synchronized via the network time protocol (NTP).

Unlike the previous systems discussed earlier [16,20,23], our system does not rely on explicit written marks (codes) since implicit correspondence between the strokes and the digital materials seemed more suited. First, pupils do not have to learn specific gestures and no error can occur because of the recognition algorithm. Second, as pupils do not control the flow of the lecture and since their attention is on the lecture, they may not have time or attention to dedicate to drawing explicit encoding marks. Finally, pupils can still write marks if they wish, using their own personal conventions. These marks will act as visual markers in their notebook (e.g., for highlighting an important aspect or for indicating a comprehension problem). Thanks to temporal associations, clicking on these marks will provide access to what the students expect. Hence, in most cases, there is no need for the system to understand the semantics of the user's marks and this would bring undesirable constraints such as forcing the pupil to use a predefined vocabulary of gestures.

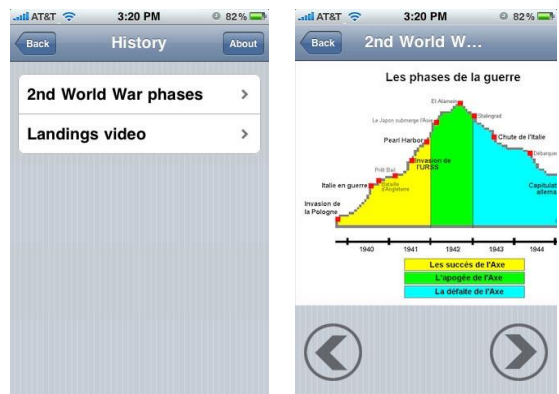

**Fig. 8.** Screenshots of U-Move - The U-Move calendar (left); Digital documents associated to a specific lesson (middle); A specific slide displayed on the mobile device (right).

**U-Move.** U-Move is a web application for mobile devices. We identified two main situations where a mobile application is useful. The first one is a fully mobile situation, as when the pupil is using public transportation, where the pupil wants to look at the documents related to the lessons but does not have access to a PC, and manipulating his or her notebook may be somewhat cumbersome. The second situation is a less mobile situation, for example, when working in a library, in which the pupil does not have access to their own PC, but still wants to look at the lecture documents.

U-Move allows the pupil to browse the lectures documents. It consists of a calendar, which is synchronized with the pupil's schedule (Fig. 8, left). When the pupil taps on a day, the application displays the corresponding lectures. By tapping on a specific lecture, a list of the documents the teacher used that day is provided (Fig. 8, middle). These can then be opened by selecting them from the list; in which case the application downloads them from the central server and displays them (Fig. 8, right). The U-Move application has been developed as a Javascript web application based on the JQuery library [6] and the JQTouch plug-in [5]. We chose to develop this tool as a web application because this solution only requires mobile web access and can work on a variety of mobile devices, regardless of their operating system.

### 5.3 Extending the Lectures Through the Notebook

While doing homework or studying lessons, the pupil will sometimes need to search for additional information on the web or other pedagogical resources. When useful information is found, these can be kept and paste into the notebook to make a link between the new document and the lesson.

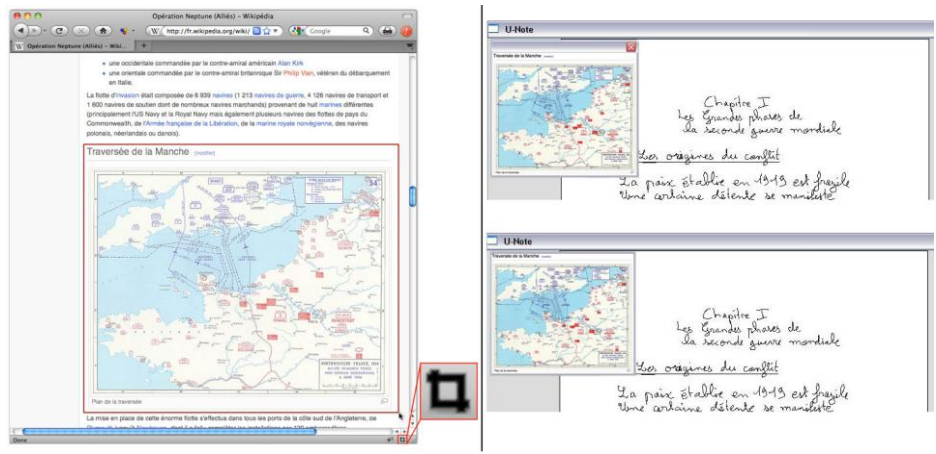

**Fig. 9.** Left: a pupil selects a region of a webpage that they want to stick to their notebook. The bottom right button is clicked, causing a bounding box to appear on the page. Right: the selected part can then be stuck to a page of the notebook.

**Adding Digital Extracts to the Notebook.** We developed a tool that allows adding pieces of documents in the digital notebook (Fig. 9). First, we developed a Firefox extension that allows one to capture a web page excerpt. This excerpt is a facsimile corresponding to the relevant subpart of the web page. The user interactively creates them by performing a drag selection (Fig.9, left). These excerpts remain attached to the original documents and can be refreshed and clicked as explained below. The U-Study module retrieves the document extracts sent by the capture tools through a socket. These excerpts appear as post-it windows (Fig. 9, right). An interesting feature is that they remain active so that the pupil can still click on the links and, for instance, view embedded files such as videos. A post-it can then be "stuck" to a given page of the notebook. Once stuck, the post-it is not active and cannot be resized to prevent unwanted modifications. It can be reactivated at will by unsticking it. The original related page can be opened in a web browser, so that these post-its are essentially bookmarks in the digital notebook.

**Adding Physical Excerpts to the Notebook.** We developed a tool that allows the pupil to print a physical interactive preview of any document opened on their PC. The pupil navigates in the U-Study menu to select the desired document to print. U-Study prints this document on Anoto paper and stores the mapping between it and the Anoto coordinates. The pupil can then cut and paste any part of the paper version of the document back into the notebook. The digital version may be opened by tapping on the piece of paper with the digital pen.

## 6 Conclusion

With U-Note we focused on helping pupils access the digital materials presented during classes. Preliminary interviews and questionnaires showed that while paper is still widely used, teachers are also increasingly using multimedia content. Consequently, we proposed to augment the pupil's notebook so that it can serve as a central medium for referencing and accessing digital information. The notebook provides a simple means of accessing digital media presented in the class, together with oral explanations and the writings on the whiteboard. Moreover, the pupils can also enrich it by creating links to their own digital documents. U-Note allows fine-grained mapping between the notes and digital media and makes it possible to access them in various situations.

Future work includes a longitudinal study with teachers and pupils. We also plan to enhance the capabilities of the system, mainly for making it possible to capture more types of digital media and to easily create digital extracts from these files at precisely defined spatial or temporal locations. Finally, security aspects and access rights are also a topic we would like to address in future versions of the system.

## 7 Acknowledgments

We gratefully acknowledge the financial support of the Cap Digital ENEIDE project that was funded by Région Ile-de-France and DGE. We thank the anonymous reviewers of this article for their relevant recommendations.